\documentclass[twocolumn,plb,aps,groupedaddress,nopacs,footinbib,preprintnumbers,floats,floatfix]{revtex4}
\usepackage{graphicx}
\usepackage{color}
\usepackage{verbatim}
\usepackage{mathrsfs}
\usepackage{amssymb}
\usepackage{amsmath}
\usepackage{bm}

\newcommand{\beq}{\begin{equation}}
\newcommand{\eeq}{\end{equation}}
\newcommand{\bea}{\begin{eqnarray}}
\newcommand{\eea}{\end{eqnarray}}

\def\gtorder{\mathrel{\raise.3ex\hbox{$>$}\mkern-14mu
             \lower0.6ex\hbox{$\sim$}}}

\def\zo{z_{\rm{obs}}}

\usepackage{booktabs}

\begin{document}

\title{Observational Constraints on Redshift Remapping}
\author{Bruce A. Bassett$^{1,2,3}$, Yabebal Fantaye$^{4}$, Ren\'ee Hlo\v{z}ek$^5$, Cristiano Sabiu$^{6}$ and Mat Smith$^{7,8}$    \\
\it $^1$ African Institute for Mathematical Sciences, 6-8 Melrose Road, Muizenberg, Cape Town, South Africa\\
\it $^2$ South African Astronomical Observatory, Observatory, Cape Town, South Africa\\
\it $^3$ Department of Maths and Applied Maths, University of Cape Town, Rondebosch 7701, South Africa\\
\it $^4$ Department of Mathematics, University of Rome Tor Vergata, Italy\\
\it $^5$ Department of Astrophysical Sciences, Peyton Hall, 4 Ivy Lane, Princeton, NJ 08544, USA\\
\it $^6$ Korea Astronomy and Space Science Institute, Yuseong-gu, Daejeon, 305-348, Korea\\
\it $^7$ Department of Physics, University of the Western Cape, Cape Town, 7535, South Africa\\
\it $^8$ School of Physics and Astronomy, University of Southampton, Southampton, SO17 1BJ, UK}

\begin{abstract}
There are two redshifts in cosmology: $\zo$, the observed redshift computed via spectral lines, and the model redshift, $z$, defined by the effective FLRW scale factor. In general these do not coincide. We place observational constraints on the allowed distortions of $z$ away from $\zo$ - a possibility we dub {\em redshift remapping}. Remapping is degenerate with cosmic dynamics for either $d_L(z)$ or $H(z)$ observations alone: for example, the simple remapping $z = \alpha_1 \zo+\alpha_2\zo^2$ allows a decelerating Einstein de Sitter universe to fit the observed supernova Hubble diagram as successfully as $\Lambda$CDM, highlighting that supernova data alone cannot prove that the universe is accelerating. We show, however that redshift remapping leads to apparent violations of cosmic distance duality that can be used to detect its presence even when neither a specific theory of gravity nor the Copernican Principle are assumed.  Combining current data sets favours acceleration but does not yet rule out redshift remapping as an alternative to dark energy. Future surveys, however, will provide exquisite constraints on remapping and any models -- such as backreaction -- that predict it. 
\end{abstract}

\maketitle

The discovery of the anomalies ascribed to dark energy has pushed cosmology into a situation where the famous quote by Bohr seems apt: {\em ``We are all agreed that your theory is crazy. The question that divides us is whether it is crazy enough to have a chance of being correct"} \cite{bohr}. The remarkable success of the $\Lambda$CDM model in matching cosmological observations is something of a Pyrrhic victory: we lack a non-anthropic \cite{weinberg} understanding of its key features. Faced with the extraordinary possibility of a non-zero $\Lambda$, the cosmology community searched for more mundane explanations (e.g. \cite{dust, subir}) and when these failed it considered more exotic alternatives such as axion-photon mixing to dim the supernovae \cite{axionphoton},  backreaction and violations of the Copernican Principle, among others. However, there is one aspect of our models that has received surprisingly little attention in the quest to challenge $\Lambda$: redshift. 

We use observational data to learn about the parameters of a cosmological theory by comparing the theory with the data at the same redshift. But what do we mean by `same'? Complex theories can have more than one redshift and even in General Relativity we are fitting an {\em effective} FLRW universe, with some average scale factor, $a_{\rm{RW}}$ to the real, perturbed, universe \cite{fitting}, much like one can fit a perfect sphere to the surface of the earth. Why should the redshift, $\zo$,  measured through the ratio of spectral line wavelengths, coincide with the redshift given by the standard model expression $1 + z = a_0/a_{\rm{RW}}$, where $a_0$ is the current value of the scale factor?  One circumstantial, but compelling, reason for this belief is the accuracy with which we compute the temperature anisotropies in the cosmic microwave background which are essentially the anisotropies in the redshifts of the photons gathered as they propagated from the surface of last scattering to us.

Nevertheless, even within pure General Relativity, $z \neq \zo$ in general; an effect we refer to as {\em redshift remapping}. Consider the definition of $\zo$ via the equation $1 + \zo = (u^a k_a)_e/(u^a k_a)_r = \nu_e/\nu_r$, where $u^a, k^a$ are the cosmological four-velocity and photon momentum tangent to the null geodesics, respectively, $\nu$ is the photon frequency and $e,r$ denote emission and reception respectively. Splitting $k^a = \nu (e^a + u^a)$ where $e^a$ is the spatial part of the photon direction ($e^a u_a = 0$ and $e^a e_a = 1$), the redshift of light can be integrated along a null geodesic parametrised by affine parameter, $\gamma$, as (see e.g. \cite{syksy}): 
\begin{equation}
1+\zo = \exp\left( \int  
\left[\frac{1}{3}\theta + \sigma_{ab} e^a e^b + \dot{u}_a e^a\right] d\gamma\right)
\label{redshift}
\end{equation}
where $\theta = u^a_{;a}$ is the expansion, $\sigma_{ab}$ is the shear and $\dot{u}_a$ is the acceleration of observers with four-velocity $u_a$. The nonlinearity of this formula implies that in general, the observed redshift, $\zo$, will not coincide with the redshift, $z$, computed via $1 + z = a_0/a_{\rm{RW}}$, after spatial averaging. 

\begin{figure}[ht!]
\centering
\includegraphics[width=\linewidth]{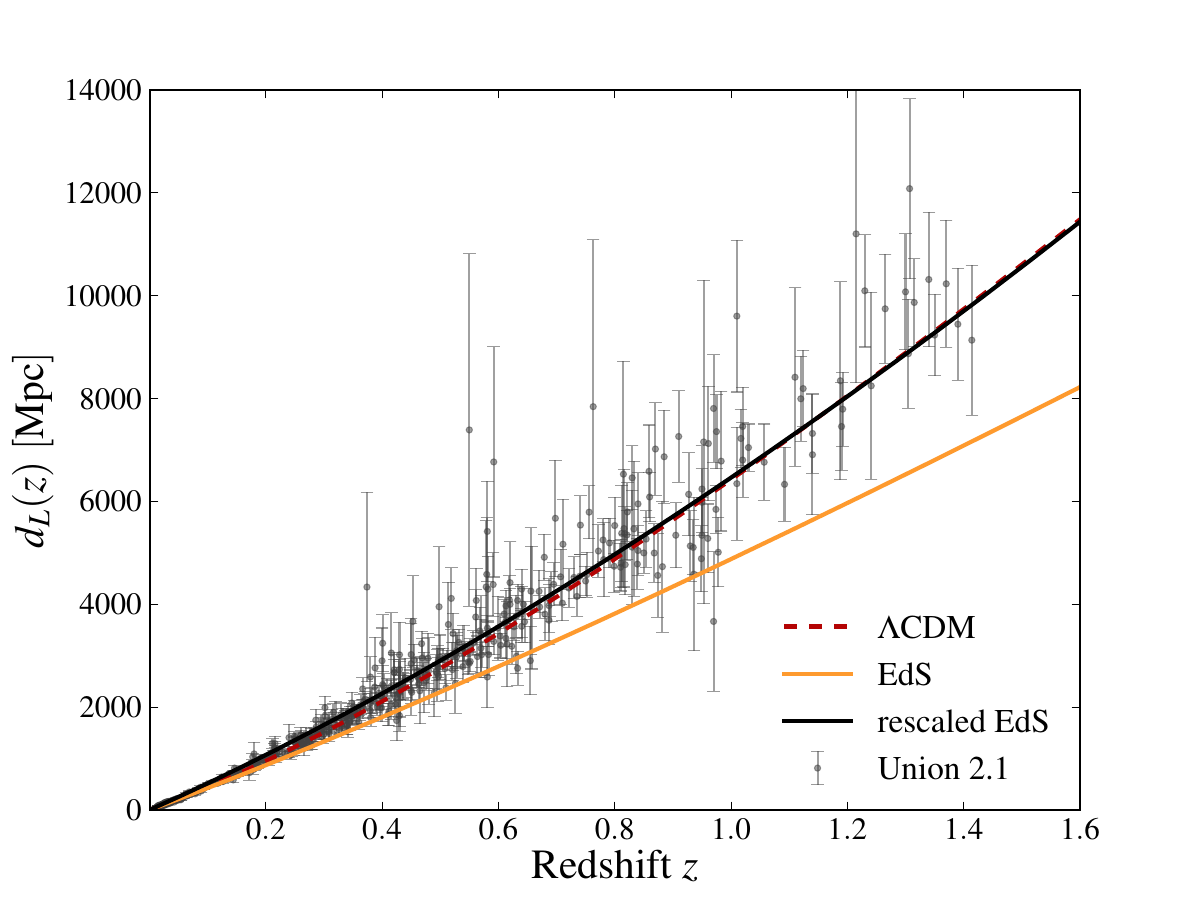}
\caption{Fitting the supernova data by increasing $z$ at fixed $d_L$. The lower, beige curve for Einstein de Sitter  (EdS: flat, $\Omega_m = 1$) is ruled out by the Union2.1 data. However, the remapped EdS model with $z = \alpha_1 \zo + \alpha_2 \zo^2$ (black solid line)  is almost indistinguishable from the concordance $\Lambda$CDM model ($\Omega_m = 0.3$, dashed line). An even simpler remapping of $z = \zo^{1.13}$, matches the concordance $d_L(z)$ but with $H_0 = 55~{\rm km s^{-1} Mpc^{-1}}$. It is clear that from supernovae alone one can neither prove our universe is accelerating nor rule out significant redshift remapping.} 
\label{dL}
\end{figure}
\begin{figure}[ht!]
\centering
\includegraphics[width=\linewidth]{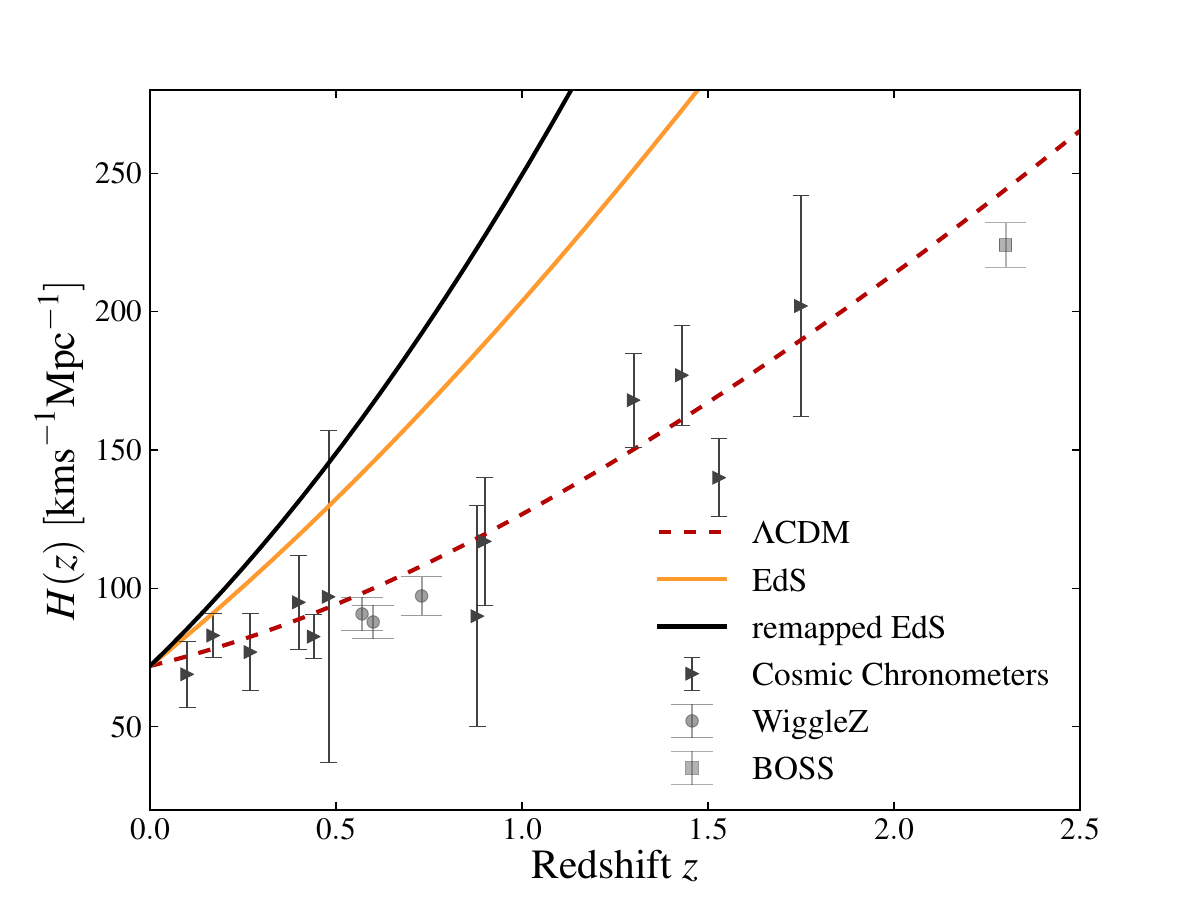}
\caption{$H(z)$ for the standard $\Lambda$CDM (bottom curve), standard EdS and the remapped EdS that fits the SNIa data in Fig.(1), (top curve). To bring the EdS model into agreement with the $H(z)$ data would require $z< z_{obs}$, the exact opposite of what is required to match the $d_L(z)$ data showing the power of combining $d_L(z)$ and $H(z)$ data. The data shown is without the changes described in the text and in section (\ref{remap}).
} 
\label{hub}
\end{figure}

For linear fluctuations around FLRW, these differences are argued to be small \cite{syksy}. Nevertheless, significant differences between $z$ and $\zo$ have been found in several studies of models which are FLRW on average but with significant inhomogeneities \cite{shrink, backreact, CF, drwho, timescape}. In such models there can even be exotic features such as a multi-valued angular diameter distance  redshift relation, $d_A(z)$ (see Fig. 5 of \cite{shrink}), caused by the deviation of $\zo$ from $z$. Physically this corresponds to a situation where the gravitational and cosmological redshifts arising from crossing an inhomogeneous region cancel. 

We emphasise that there are other effects that can also yield $z \neq \zo$. Consider bi-metric theories in which photons couple to a metric, $\tilde{g}_{ab}$, which is different than the metric, $g_{ab}$, to which matter couples; see e.g. \cite{vsl}. Since radiation is today sub-dominant we want to infer the properties of the matter metric since this will allow us to deduce the dominant constituents of the universe. The frequency of light is determined by the null geodesic equation $k^a \tilde{\nabla}_a k^0 = 0$ where $\tilde{\nabla}_a$ is the covariant derivative w.r.t $\tilde{g}_{ab}$, which implies $z \neq \zo$ if $\tilde{g}_{ab} \neq g_{ab}$. Explicit examples of this are provided by Born-Infeld and Euler-Heisenberg nonlinear electrodynamics, which can be formulated as light propagating on a different effective metric, which modifies the standard redshift predictions \cite{nled}. 

In the same vein, it has long been known  that photon propagation in 1-loop QED in a curved background is superluminal at low-frequencies \cite{drumhath}, with similar behaviour in the Casimir vacuum, known as the `Scharnhorst' effect \cite{scharn}. Similarly electromagnetic radiation scatters off spacetime curvature creating tails \cite{tails} that travel inside the null cone, not just on it, even classically. These do not threaten causality \cite{superlum1,scharn1} but they suggest that redshift remapping may be generic, though perhaps a small effect 
\footnote{One can argue that generically the frequency will be unchanged but the wavelength will change if the speed of light changes, as occurs in standard media with refractive indices different from unity. Spectrographs used to measure $\zo$ exploit diffraction and interference and hence respond to the wavelength. Hence variations in the speed of light should induce redshift fluctuations.}. 

Two final examples of remapping are provided by conformally FLRW models \cite{matt} and the non-metric generalisation of the Pleba\'{n}ski formulation of General Relativity \cite{nonmetric1}, which leads to redshift remapping depending on the ambient spatial curvature. 
 
We are not, of course, pushing any of these models or effects as likely, rather they suggest that the relation $1 + \zo = a_0/a_{\rm{RW}}$ should not be taken as God-given. Rather the relation between $z$ and $\zo$ should either be rigorously derived or tested experimentally. What observational limits, then, can one place on deviations of $z$ from $\zo$? 
 
{\em Observational Implications --} Interpreting cosmological observations becomes significantly more complex in the presence of remapping. We now fit data at an observed redshift, $\zo$, to a theory with unknown parameters at an unknown model redshift, $z$. Unless the theory predicts $z$ given $\zo$,  parameter constraints weaken due to marginalising over the unknown model redshift $z$. 

Redshift remapping will in general depend on environment. Here we focus on generic effects implied by remapping in which the model redshift, $z$, is only a function of  $\zo$, $z = f(\zo)$. In particular, we study the model:
\beq
z = \alpha_1\,\zo + \alpha_2\,\zo^2\,.
\label{basiceq}
\eeq 

One result of remapping is that it changes the values of derived data points based on the dynamics of matter. Consider Baryon Acoustic Oscillations (BAO); e.g. \cite{baoreview}. The transverse BAO scale is relatively unaffected: it still gives the angular-diameter distance, $d_A(z)$, but now at redshift $z=f(\zo)$ instead of $\zo$. The radial BAO scale however is given by $\Delta z/H(z)$, but $\Delta z$ is not observable: $\zo$ is. Therefore the BAO scale becomes $z'~\Delta \zo/H(z)$ where $z' = dz/d\zo$. This changes the derived value of $H(z)$, a novel feature in these theories. In addition we must scale all BAO results by the size of the sound horizon, $r_s \sim (\Omega_M h^2)^{-1/4}$. 
This change is important, particularly for $H(z)$; we show its impact on current constraints in Figures (\ref{nos}) and (\ref{nosa1a2}).  The changes in any specific model of remapping may be even more complicated. For example, in the non-metric version of modified General Relativity \cite{nonmetric2}, perturbations evolve with a time-dependent effective speed of sound that does not vanish even during matter domination. 

The predictions of redshift remapping for any specific theory must be computed from first principles but we can nevertheless make significant progress for theories given by Eq. (\ref{basiceq}). We consider two limiting cases: the first ({\em Case A}) assumes that redshifts are altered but distances are unchanged and hence are still linked to the standard FLRW parameters through $H(z)$ and the Friedman equation. In the second and much more general case ({\em Case B}) we assume that everything is altered and assume only that the distance duality relation holds in the underlying theory: $d_L(z) = (1+z)^2 d_A(z)$. Here we cannot assume General Relativity and thus there is no link between the Hubble rate, $H(z)$ and distances \footnote{If we assume FLRW then we recover the usual link between $H(z)$ and $d_L(z)$ since this does not assume General Relativity \cite{cop1}. Since we want to be as general as possible we do not consider this further but note that the constraints will interpolate between those of {\em Case A} and {\em Case B}.}.   

Let us begin with {\em Case A}. If $H(z)$ is monotonic then the luminosity distance $d_L(z)$ is a monotonically increasing function of $z$. This means that one can map the $d_L(z)$ of {\em any} such FLRW cosmology into that of any other FLRW cosmology, by suitable choice of Eq. (\ref{basiceq}). Instead of increasing distances at fixed $z$ by increasing $\Omega_{\Lambda}$, one can instead increase $z$ at fixed $\zo$ and $d_L$. 

\begin{figure}[ht!]
\centering
\includegraphics[scale=.42]{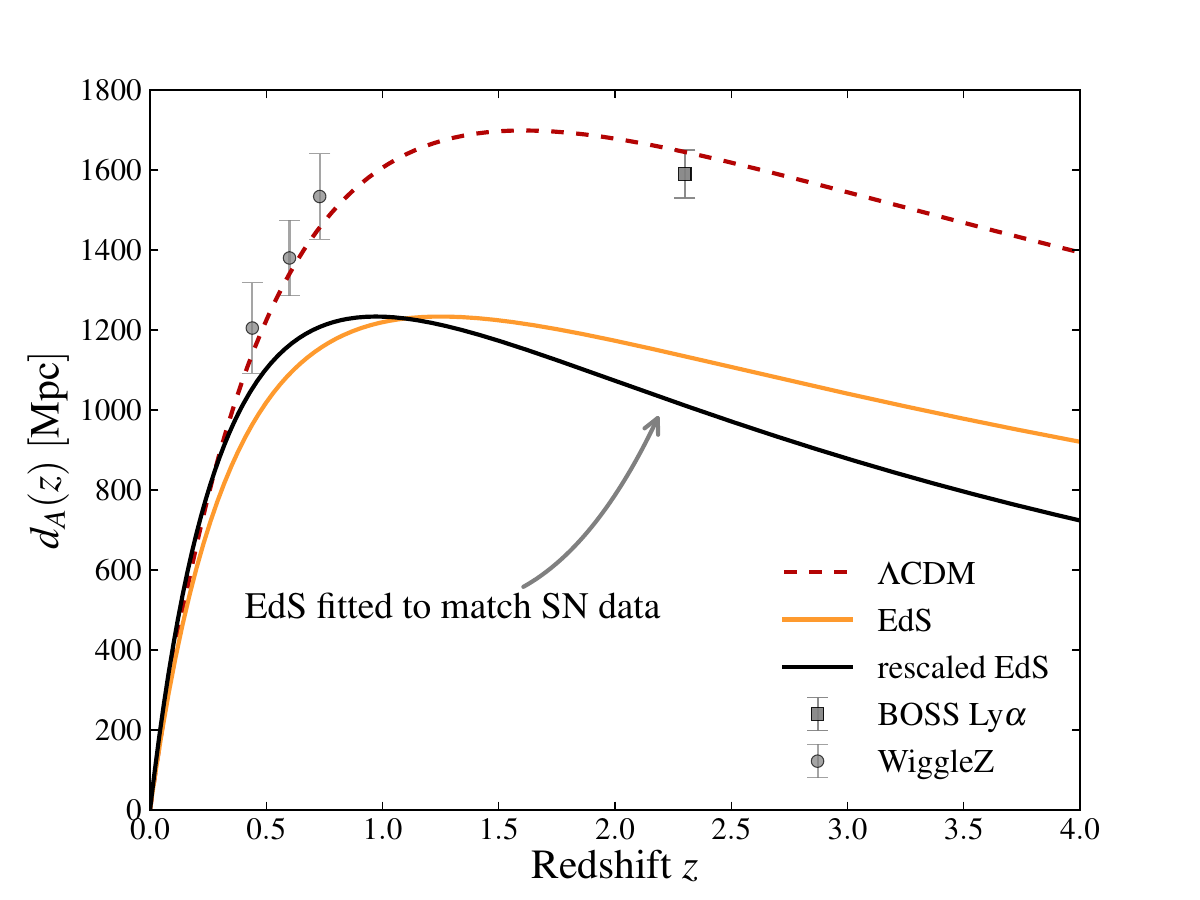}
\caption{The angular diameter distance, $d_A(z)$, for the same three models as in Fig.(\ref{dL}). Since $d_A = d_L/(1+z)^2$ and $d_A$ has a peak, the remapped EdS model is an even worse fit to concordance data than pure EdS, leading to an apparent violation of distance duality, independent of any model assumptions. We plot the BOSS Lyman-$\alpha$ data point at $z=2.3$. We emphasise, however, that for many models leading to $z$-remapping, the data themselves need to be consistently re-analysed.} 
\label{dA}
\end{figure}

Interestingly, Eq. (\ref{basiceq}) allows us to map the $d_L(z)$ of the flat, decelerating Einstein-de Sitter (EdS) universe ($\Omega_m = 1$) almost exactly onto that of the concordance $\Lambda$CDM cosmology if we assume that distances are unaffected by remapping. Two example mappings are $z = \zo^{1.13}$, with $H_0 \simeq 55~{\rm km\, s^{-1}Mpc^{-1}}$ and $z = 1.2~\zo + 0.089 ~\zo^2 $ with $H_0 = 72{\rm ~km\,s^{-1} Mpc^{-1}}$. The latter gives the black curve in Fig.(\ref{dL}) which is almost indistinguishable from the concordance $\Lambda$CDM (dashed curve). In Fig.(\ref{dlhub}) we show the $z=f(z_{obs})$ mapping required to make the EdS $d_L(z)$ {\em exactly} equal to the $\Lambda$CDM $d_L(z)$ for all redshifts.  This makes remapping one of only two models that fit major cosmic data as well as $\Lambda$CDM with the same number of free parameters, the other being non-local gravity which goes further and fits all current data as well as $\Lambda$CDM \cite{NLG}.

\begin{figure}[ht!]
\centering
\includegraphics[width=\linewidth,height=7.5cm]{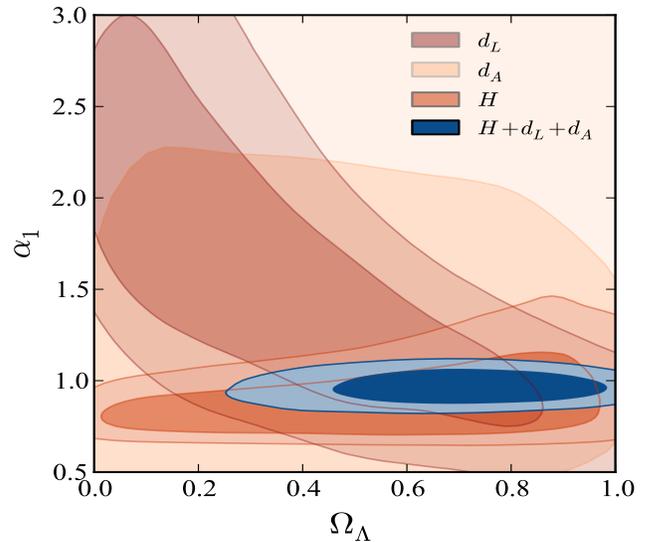}
\caption{Marginalised 1 \& 2-$\sigma$ contours for current $d_L(z)$, $H(z)$ and $d_A(z)$ data from Union2.1, WiggleZ, BOSS and cosmic chronometers for the model $z = \alpha_1 \zo + \alpha_2 \zo^2$. While each dataset alone is consistent with $\Omega_{\Lambda}=0$, the combined datasets confirm acceleration ($\Omega_{\Lambda} = 0.68 \pm 0.15$), strongly constrain redshift remapping ($\alpha_1 = 0.97 \pm 0.06$) and are consistent with no remapping $(\alpha_1,\alpha_2) = (1,0)$. Note the data have been rescaled as per the discussion in the text, though this does not change our main conclusions, see Fig.(\ref{nos})} 
\label{MCMC}
\end{figure}

For either $d_L(z)$ or $H(z)$ data alone there is a perfect degeneracy between $\Omega_{\Lambda}$ and remapping, implying that a decelerating universe is actually perfectly consistent with the supernova data. In fact, for a remapping function of the form $z = \alpha_1\,\zo + \alpha_2\,\zo^2$, the likelihood for  $\Omega_{\Lambda}$ with the Union2.1 supernova data \cite{union2} peaks around $0.15$ and is fully consistent with $\Omega_{\Lambda} = 0$, as shown in Fig.(\ref{1dlike}).          

However, combined measurements of both $d_L(z)$ and $H(z)$ break the degeneracy. A model which fits the supernova data with an EdS model as in Fig. (\ref{dL}) has a significantly larger $H(z)$ at any redshift than in the concordance model  with the same $H_0$ and $\Omega_K$; as shown in Fig. (\ref{hub}), breaking the degeneracy between remapping and $\Omega_{\Lambda}$. This comes from two places: the EdS $H(z)$ is larger at the same $z_{obs}$, {\em and} we have to compare the model to the data at $z > z_{obs}$, and since $H(z)$ is increasing with $z$, the difference is even larger \footnote{It is possible to partially circumvent this argument, however: by playing with $H_0$ and $\Omega_K$ in the two models, it is  this argument. If this were the case, one would expect there to be a mismatch between the cosmological fit value of $H_0$ (assuming no redshift remapping) and local measurements, perhaps not dissimilar to the current tension \cite{boss2,SH03}.} 

By themselves, simultaneous measurements of $d_L(z)$ and $H(z)$ do not help constrain remapping at all in {\em Case B} however, since there is no link between them in general. Fortunately we can break the degeneracies caused by redshift remapping another way: in any metric theory of gravity where photon number is conserved, distance duality must hold \cite{dd}, namely $d_A(z) = d_L(z)/(1+z)^2$. Whereas $d_L(z)$ grows without bound, $d_A(z)$ has a peak around $z \sim 1.5$ corresponding to the minimum apparent size of an object. This implies that we {\em cannot} match $d_A(z)$ curves in any FLRW model by redshift remapping, as we could for $d_L(z)$. Since acceleration truly increases $d_A(z)$, there is no redshift in an EdS model with the same $H_0$ and $\Omega_K$ that will match the maximum distance in the $\Lambda$CDM model. If redshift remapping is active ($z \neq \zo$), it can therefore be detected as an apparent violation of distance duality, as shown in Fig.(\ref{dA}). In particular, for redshifts above the maximum in $d_A(z)$,  remapping with $z > \zo$ (as required to fit the larger $d_L(z)$ distances) leads to a {\em decrease} in the predicted $d_A(z)$, irrespective of the choice of $H_0$, another distinctive prediction of remapping. 

This effect is illustrated in Fig. (\ref{dA}) which shows the remapped EdS model which mimics the $\Lambda$CDM $d_L(z)$ shown in Fig. (\ref{dL}). It has a lower $d_A$ than even the unremapped EdS model. For comparison we show the BOSS survey data from the BAO  in the Lyman-$\alpha$ forest at $z \sim 2.3$ \cite{boss} that is consistent with the concordance $\Lambda$CDM model but strongly in disagreement with the remapped EdS model which fits the supernova data. One must caveat that the error distributions for data such as this are highly non-Gaussian so caution should be taken in using them far from the fiducial value \cite{sting,boss}. 

In Figs (\ref{MCMC}) and (\ref{a1a2}) we show the joint MCMC analysis of current $d_L, H$ and $d_A$ data from Union2.1 \cite{union2}, BOSS \cite{boss,boss2}, WiggleZ \cite{wigglez} and cosmic chronometers \cite{cc} for {\em Case A}. This yields the constraint $\Omega_{\Lambda} = 0.68 \pm 0.15$ after marginalising over $\alpha_1, \alpha_2$ and $H_0$, showing that redshift remapping is not an alternative to cosmic acceleration if we assume both that General Relativity holds and that distances are unaffected by remapping. Fig.(\ref{1dlike}) shows this also from the point of view of the 1-d likelihoods. Nevertheless, allowing for remapping does increases errorbars by $50-100\%$, relative to the no-remapping case, as shown in Table (\ref{tab1}).  
 
These assumptions are restrictive, however. If remapping is active it is reasonable to expect that distances are altered too and that gravity may be significantly different from General Relativity. In this situation ({\em Case B}) there is now no link between $H(z)$ and distances and no equivalent of the Friedmann equation, so we do not, {\em a priori}, have parameters like $\Omega_{\Lambda}$. Despite this generality it turns out we can make powerful progress if we make one assumption:  that distance duality is preserved at the fundamental level, i.e. $d_L(z) = (1+z)^2 d_A(z)$ \footnote{Without this assumption it is extremely difficult to say anything about remapping since $d_L(z)$, $d_A(z)$ and $H(z)$ are all independent.}

As discussed above, redshift remapping will lead to apparent violations of distance duality when phrased in terms of $\zo$.  With good $d_A(\zo)$ and $d_L(\zo)$ data at the same observed redshifts we can immediately infer the $z=f(\zo)$ remapping via: 
\beq 
z(\zo) = \left(\frac{d_L(\zo)}{d_A(\zo)}\right)^{1/2} - 1
\label{DD}
\eeq
Note that this does {\em not} assume that distances are unchanged: they can be arbitrarily deformed, so long as distance duality is preserved in terms of the theoretical redshift $z$. This allows us to unambiguously disentangle remapping from any physics which changes gravity or dynamics. This will allow future data to provide excellent constraints on remapping as shown by the darker contours in Fig. (\ref{noassume1}), which is one of the main results of this paper.   

Future supernova and lensing measurements \cite{lensing} will yield $d_L$ and $d_A$ to high accuracy but to get $H(z)$ we require BAO, which depends linearly on the unknown sound horizon scale, $r_s$, which can differ significantly from its $\Lambda$CDM value in general. This creates a new degeneracy, but one that can be largely broken by measuring distance duality at multiple redshifts. Alternatively, a more general way of breaking the degeneracy is to compare the BAO-derived $H(z)$ to an alternative, $r_s$-free measurement of $H(z)$, such as from cosmic chronometer measurements based on measuring the ages of cosmic objects \cite{cc}. Demanding that the two measurements of $H(z)$ agree at the same redshifts fixes $r_s$ and breaks the degeneracy between remapping and $r_s$. 

\begin{figure}[ht!]
\centering
\includegraphics[width=9.2cm,height=6.2cm]{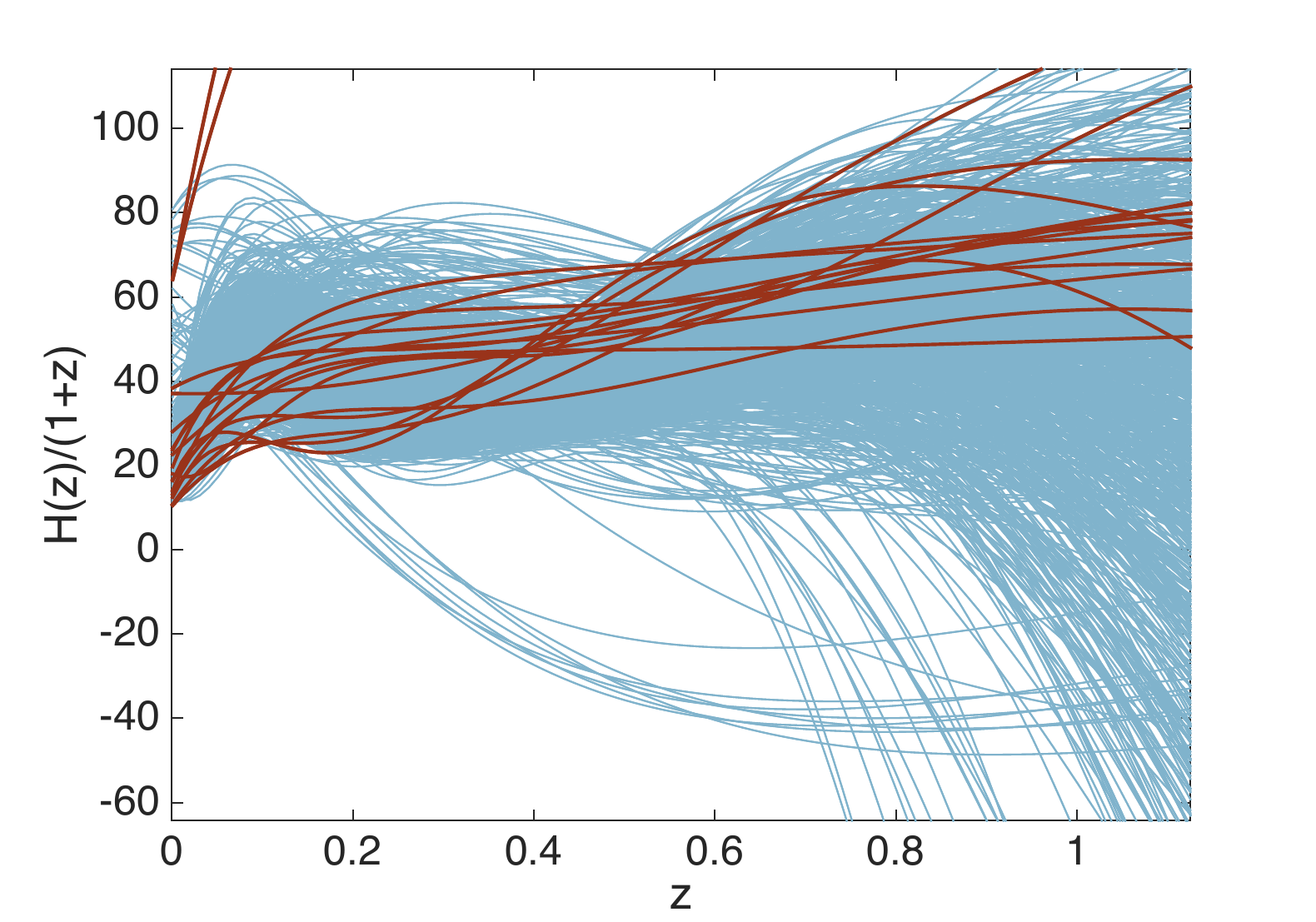}
\caption{Acceleration is still favoured: evolution of $\dot{a} = H(z)/(1+z)$ for 1000 randomly chosen MCMC points which fit the 13 free parameters $\{\alpha_1, \alpha_2, H_0, \eta_n, \delta_n \}$ from the {\em Case B} expansions of $d_L(z), d_A(z), H(z)$ given by Eq's (\ref{expand}) to all current data.  Approximately $99\%$ of all the MCMC curves exhibit acceleration (decreasing $H(z)/(1+z)$ with increasing $z$) somewhere in the range $0.2 < z < 0.8$ as illustrated by the light (blue) curves which dominate over the few decelerating (darker red) curves, several of which in fact show acceleration outside the range $0.2 < z < 0.8$. This implies that cosmic acceleration is preferred over remapping as an explanation of current data even in the most general case. Interestingly, there are more examples of collapsing phases where $\dot{a} < 0$, than deceleration. Since we are using a general Taylor series expansion of $H(z)$ rather than the Einstein equations these collapsing phases are to be expected in regions where there is not much data.} 
\label{dota1}
\end{figure}

We apply this method now both to current and future data using the cosmic chronometer measurements of $H(z)$ to break the BAO $H(z)$ degeneracy with $r_s$. Since we can neither assume General Relativity nor FLRW we use 5th-order Taylor series expansions in both the distances and $H(z)$ in powers of $\epsilon \equiv z/(1+z)$, which are well-behaved at $z > 1$ and provide excellent fits to $\Lambda$CDM simulations: 
\bea
d_A(\epsilon)&=& \sum_{n=1}^5 \delta_n \epsilon^n \nonumber \\
H(\epsilon)&=& H_0\left [1 + \sum_{n=1}^5 \eta_n \epsilon^n \right] 
\label{expand}
\eea
The coefficients $\delta_n, \eta_n, H_0$ are unknown and must be fit simultaneously along with $\alpha_1,\alpha_2$ and then marginalised over. Once a specific model of gravity, or FLRW geometry is assumed there will be tight links between these parameters (as in the case of General Relativity) but we do not assuming any such model here. Redshift remapping deforms $\epsilon_o$ and hence the $\alpha$ parameters connect the fitting of distance and Hubble parameters.  

Using this approach we test for acceleration by looking for $\ddot{a} > 0 \Rightarrow  d(H(z)/(1+z))/dz < 0$  \cite{sutherland}.  We find that $\sim 99\%$ of points in our MCMC chains show acceleration somewhere in the range $0.2 < z < 0.8$ given current data, as shown in Fig. (\ref{dota1}) and discussed in more detail in appendix (\ref{caseB}). This is very similar to results from simulated data from $\Lambda$CDM with no remapping and the same current data covariance matrix.  A random selection of 1000 such curves are shown in Fig. (\ref{dota1}) of which only 18 do not have acceleration in the selected redshift range, and several of these show acceleration outside the range $0.2 < z < 0.8$.  This appears to strongly confirm cosmic acceleration and rule out remapping but when we simulate a decelerating Einstein-de Sitter universe without remapping we find that about 54\% of the chain elements spuriously exhibit acceleration in the range $0.2 < z < 0.8$ due to oscillations in the expansions in Eq. (\ref{expand}). As a result, although current data are fully consistent with an accelerating universe and no remapping, we cannot yet rule out remapping as an alternative to cosmic acceleration. 

In Fig. (\ref{noassume1}) we show the constraints on $\alpha_1, \alpha_2$ for current and future (small contours) distance and Hubble rate data. They illustrate that the data are consistent with no remapping at 2-$\sigma$.  For the future surveys we assume a $\Lambda$CDM  fiducial model with $\alpha_1 = 1, \alpha_2 = 0$ and $0.5\%$ measurements of $d_L, d_A$ and $H(z)$ from both BAO and cosmic chronometers in $\Delta z = 0.1$ bins out to $z=1.5$. The error ellipses show that future data will have exquisite sensitivity to detect redshift remapping. 
 
When we simulated an Einstein-de Sitter model with no remapping with the future data covariance matrix we find that none of the chains show acceleration (see appendix (\ref{caseB}))  showing that as cosmological data improves we will be able to completely rule out remapping as an alternative to acceleration. It is remarkable that we will be able to constrain redshift remapping so strongly even in  {\em Case B} where we make almost no assumptions about the underlying gravitational theory (other than assuming distance duality), encoded in the 13 free parameters of these expansions, or how strongly the remapping is allowed to distort distances and the expansion rate.

\begin{figure}[ht!]
\centering
\includegraphics[width= 8.5cm]{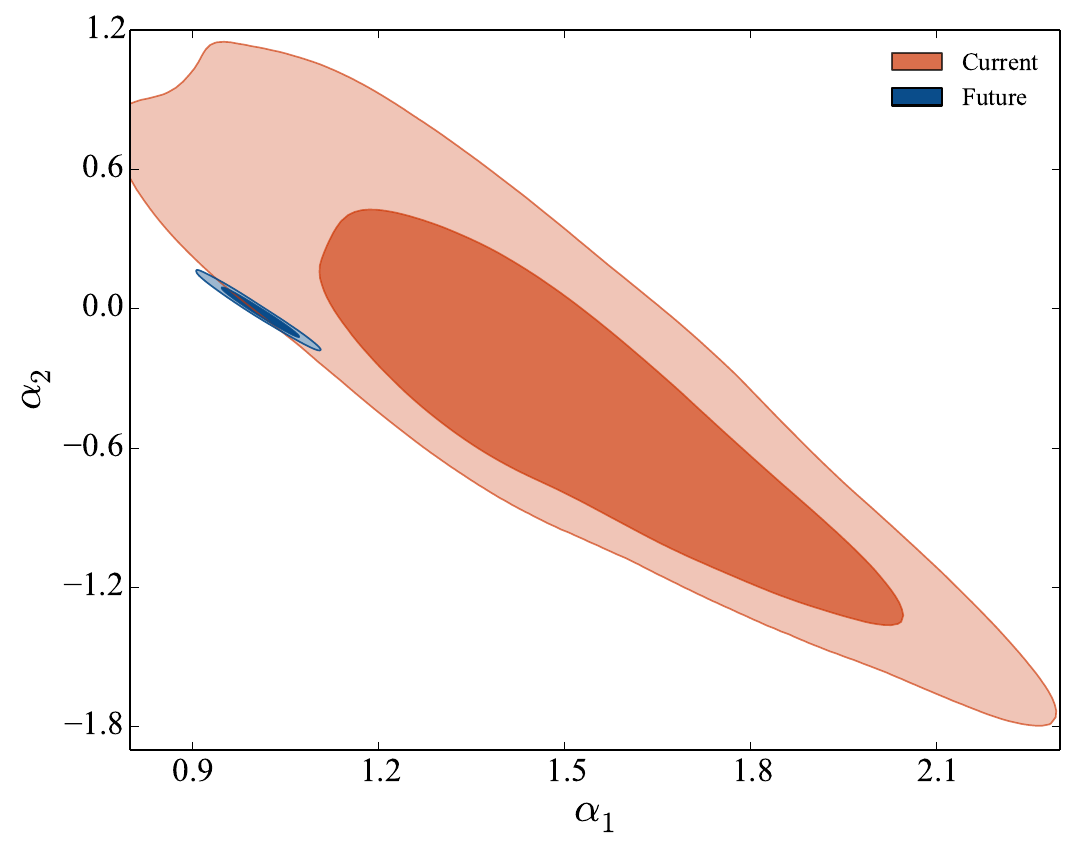}
\caption{1 and 2-$\sigma$ contours on $\alpha_1, \alpha_2$ for current (large contours) and future (small dark contours) $d_L(z)$, $H(z)$ and $d_A(z)$ data with no assumptions on how remapping affects distances and $H(z)$: the coefficients $\delta_n, \eta_n$ of Eq. (\ref{expand}) are marginalised over. We find that the data are consistent with no remapping at 2-$\sigma$. Future data will provide exquisite tests of redshift remapping which will provide a new generation of tests of exotic theories.} 
\label{noassume1}
\end{figure}

{\em Conclusions} -- In this paper we have discussed the possibility of redshift remapping in which model redshifts, $z$ and observed redshifts, $\zo$, do not coincide. Remapping has major implications: it implies that measurements of $d_L(z)$ or Hubble expansion $H(z)$ alone can never prove cosmic acceleration, no matter how good the data, if redshift remapping is not constrained separately. 

We have demonstrated, however, that remapping can be tightly constrained by combining multiple datasets which break the dynamics-remapping degeneracy. Further, if one assumes that distance duality holds fundamentally, then current data are consistent with no remapping at $2-\sigma$ and prefer acceleration over remapping as an explanation for the current data, even when no assumptions are made about how remapping affects distances and the expansion rate of the cosmos. Due to the generality of the expansions we use, the current data is not good enough to rule out remapping but we show that future data will completely exclude remapping as an alternative to acceleration.

Detection of redshift remapping - even in trace amounts - would provide powerful evidence of fundamentally new physics. As the quality of observational measurements and theoretical modeling improve, constraints on possible deviations of the observed redshift $\zo$ from the FLRW model prediction $1 + z = a_0/a_{\rm RW}$ for the redshift will improve significantly. In particular, we expect that any models based on backreaction or small-scale nonlinearities that predict significant redshift remapping will be tightly constrained or, indeed, detected through this method. 

{\em Acknowledgements} -- This paper originated as a project at the 2009 JEDI-III workshop. We thank Marco Regis, Petja Salmi, Jean Claude Kubwimana and Patrice Okouma for early work on the idea. We thank Pier-Stefano Corasaniti, Michael Feast, David Kaiser, Roy Maartens, Robert de Mello Koch, Michelle Lochner,  Stefano Liberati, Navin Sivanandam, Patricia Whitelock and especially Martin Kunz and Ignacy Sawicki for useful discussions. We thank the anonymous referee for comments that significantly enhanced the paper. BB thanks LUTH-Observatoire de Paris for hospitality and acknowledges funding from the NRF South Africa. YF is supported by ERC Grant 277742 Pascal.


\section{Additional Online Material: Case A}

\begin{figure}[ht!]
\centering
\includegraphics[scale=.42]{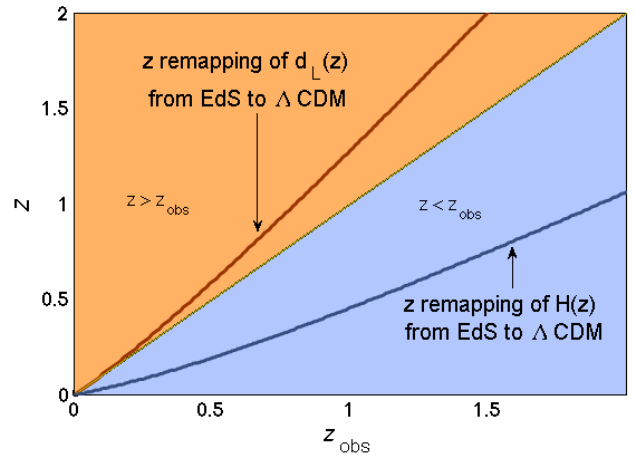}
\caption{
The $z=f(z_{obs})$ remapping functions required to exactly match standard concordance $\Lambda$CDM luminosity distance (top curve) and Hubble rate (bottom curve) with the pure Einstein-de Sitter model (flat, $\Omega_m = 1$), all with $H_0 = 72{\rm \,km\, s^{-1} Mpc^{-1}}$. For any model, increasing distances relative to the fiducial model with remapping requires $z > z_{obs}$ while decreasing the Hubble rate (usually required to get larger distances in standard cosmology) requires $z < z_{obs}$. This implies that general redshift remapping will leave very clear signatures, and allow strong constraints on remapping.  
}  
\label{dlhub}
\end{figure}

\begin{figure}[ht!]
\centering
\includegraphics[width=\linewidth]{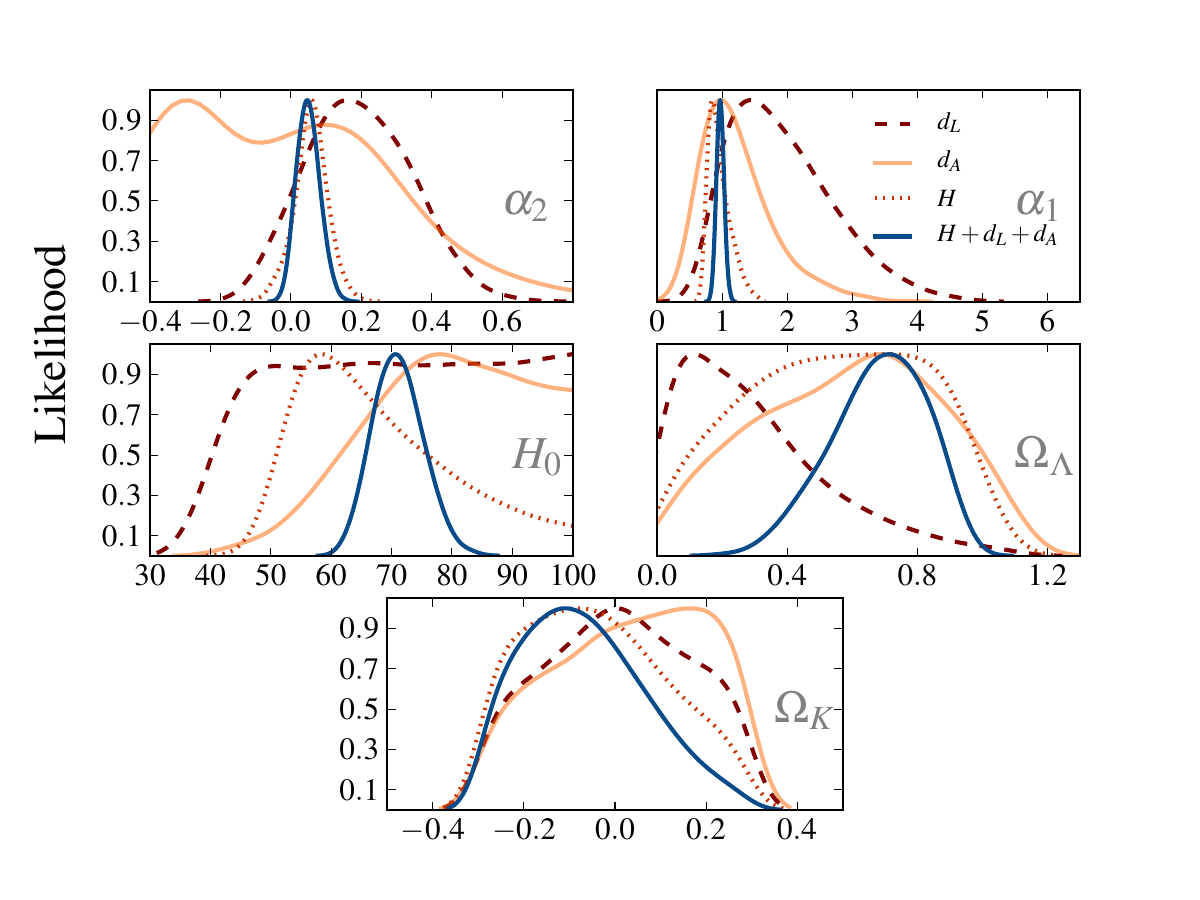}
\caption{One-dimensional likelihoods for various parameters showing the power of combining multiple different probes. Note that the Union2.1 data for $d_L(z)$ (red dashed curve) actually prefer a low value of $\Omega_{\Lambda}$  as an explanation of the supernova data, but this disappears when the $H(z)$ and $d_A(z)$ data is included.} 
\label{1dlike}
\end{figure}

Here we provide additional details for our {\em Case A} results.  The combined $d_L(z)$, $d_A(z)$ and $H(z)$ data provides powerful constraints, as shown in the 1-d likelihoods in Fig.(\ref{1dlike}). The origin of this constraining power comes from the insight that mimicking distances requires a mapping that is in tension with the remapping required to mimic $H(z)$, as shown in Fig.(\ref{dlhub}). To understand this in general consider the formula: 
\begin{equation}
d_L(z) = \frac{(1+z)}{H_0\sqrt{-\Omega_K}}\sin\left( \sqrt{-\Omega_K} \int \frac{dz}{E(z)}\right)
\end{equation}
applicable for all values of $\Omega_K$. As usual $E(z) = H(z)/H_0$ where
\begin{equation}
E(z) = \sqrt{\Omega_M(1+z)^3 + \Omega_{\Lambda}}
\end{equation}
with $\Omega_K = \Omega_M + \Omega_{\Lambda}$

In this paper we quote constraints for remappings of the form: 
\begin{equation}
z = \alpha_1\,\zo + \alpha_2\,\zo^2\,.
\label{a1a2eq}
\end{equation}
This is natural in a Taylor-series sense, and is clearly a good approximation to the curves in Fig.(\ref{dlhub}). However, other remappings such as $z = \zo^{\alpha}$ also allow good fits to the SNIa data, albeit with a lower $H_0 \simeq 55~{\rm km\, s^{-1}Mpc^{-1}}$.  

If we expand $d_L(z)$ and $H(z)$ in $\zo$ around a flat universe:
\begin{eqnarray}
d_L(z) &=& \frac{c}{H_0}\left[ \alpha_1 \zo + [\alpha_2 + \frac{\alpha_1^2}{4}(1 + 3\Omega_{\Lambda})] \zo^2  + {\cal O}(\zo^3)\right]\nonumber \\ 
H^2(z) &=& H^2_0\large[1 + 3\alpha_1(1-\Omega_{\Lambda})\zo + 3[\alpha_2 + \alpha_1^2](1-\Omega_{\Lambda})\zo^2 \nonumber\\ 
&& + \,\,{\cal O}(\zo^3)\large{]} 
\label{taylor}
\end{eqnarray}
we see that the redshift remapping parameters $\alpha_{1,2}$ are degenerate with the usual cosmic paramters $H_0, \Omega_{\Lambda}$ for both observables. However, fortunately the degeneracies are very different, suggestive that combining $d_L(z)$ and $H(z)$ will break the degeneracies. For example, if we set $\alpha_1 = 1$, then the $d_L(z)$ degeneracy is $\alpha_2 + \frac{3}{4}\Omega_{\Lambda} = const.$ while for $H^2(z)$ it is $\alpha_2 + 1 = const/(1-\Omega_{\Lambda})$. Increasing $\Omega_{\Lambda}$ leads to opposite behaviour for $\alpha_2$ in each case.  

\begin{figure}[ht!]
\centering
\includegraphics[width=\linewidth,height=7.5cm]{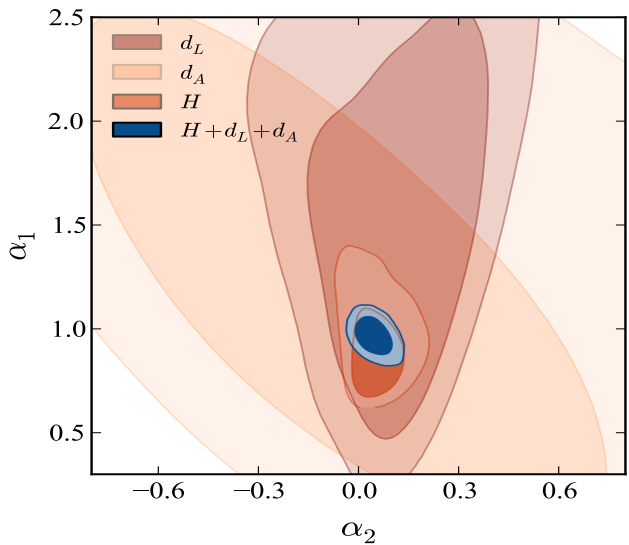}
\caption{Marginalised 1 and 2$-\sigma$ constraints on the remapping parameters in the expansion $z = \alpha_1\, \zo + \alpha_2\, \zo^2$ for each dataset alone and with all combined. $d_L(z)$ and $d_A(z)$ are each highly degenerate alone, but orthogonal, while $H(z)$ alone constrains remapping well; see Fig.(\ref{nosa1a2}). All the data together strongly constrains remapping, yielding $\alpha_1 = 0.97 \pm 0.06, \alpha_2 = 0.05 \pm 0.03$,  consistent with no redshift remapping.} 
\label{a1a2}
\end{figure}

\begin{figure}[ht!]
\centering
\includegraphics[width=\linewidth,height=7.5cm]{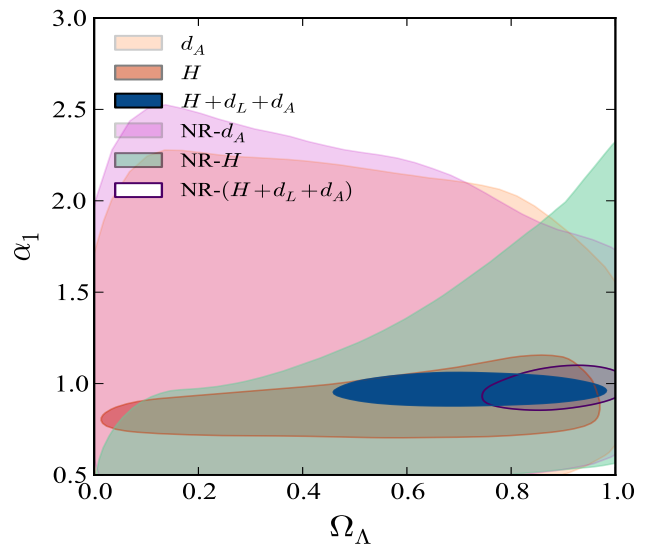}
\caption{Impact of including the effects of redshift remapping of the values of the $d_A$ and $H(z)$ data on the MCMC-derived $1-\sigma$ contours. Here ``NR" indicates the raw data (No Remapping) has been used. No prefix indicates the data have been rescaled according to Eqs (\ref{ccrm}) and (\ref{baorm}). The impact on $H(z)$ and the combined contours is significant: correct treatment of the data is important.} 
\label{nos}
\end{figure}

\begin{figure}[ht!]
\centering
\includegraphics[width=\linewidth,height=7.5cm]{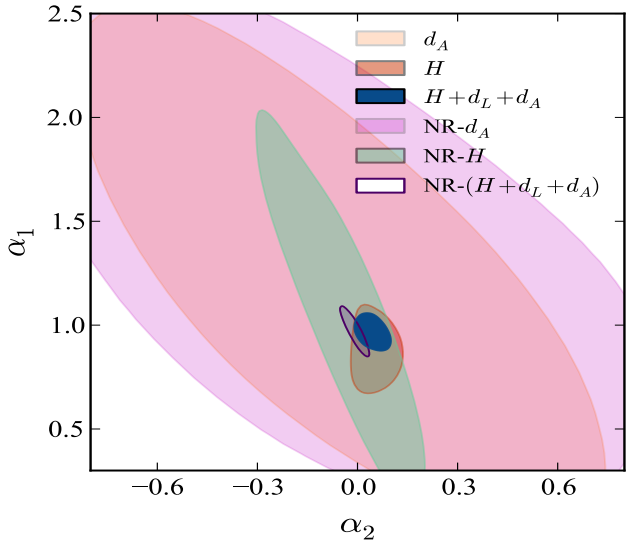}
\caption{Impact of including the effects of redshift remapping of the $d_A$ and $H(z)$ data on the MCMC-derived $1-\sigma$ $\alpha_1-\alpha_2$ contours. Here ``NR" indicates the raw data (No Remapping) has been used. No prefix indicates the data have been rescaled according to Eq's (\ref{ccrm}) and (\ref{baorm}). While the effect on $d_A$ is rather minimal, the effect on $H(z)$ is dramatic, illustrating that correct treatment of the data is important, and that multiple ways of measuring the same quantity can be powerful. Note that, despite the fact that the $H(z)$ contour shrinks when the data is remapped, after marginalisation the combined contour is actually larger after remapping, than when the data is not remapped.} 
\label{nosa1a2}
\end{figure}

We can illustrate this non-perturbatively. Imagine we have two models, $A$ and $B$, both with the same values of $H_0$ and $\Omega_K$. If $d_L^A(z) > d_L^B(z)$ at any $z$ then for redshifts where $\sqrt{-\Omega_K}\int E_A^{-1} dz < \pi/2$ (which is true up to at least $z=3$ for reasonable parameter values), then $\int E_A^{-1} dz > \int E_B^{-1} dz$ and hence, on average $E_A(z) < E_B(z)$. Hence, if one wants to mimic a model with larger distances by using redshift remapping one needs $z > \zo$ but matching $E(z)$ with remapping requires $z < \zo$. You can't have your cake and eat it. The result is a tight constraint on the redshift remapping parameters $\alpha_1, \alpha_2$, as shown in Fig.(\ref{a1a2}).

Returning to the remapping Eq. (\ref{a1a2eq}), Hubble measurements alone strongly constrain redshift remapping when the data are remapped, as discussed below. Note also that the $d_A$ and $d_L$ degeneracy directions are orthogonal in the $\alpha_1-\alpha_2$ plane, which is partly due to distance duality:
\begin{equation}
d_L(z) = (1+z)^2 d_A(z)\,.
\label{eqdd}
\end{equation}
which holds in any metric theory of gravity where photons travel on null geodesics and photon number is conserved \cite{dd}. Note that we distinguish between the closely-related reciprocity relation and distance duality because the reciprocity relation is pure differential geometry. It only becomes Eq. (\ref{eqdd}) with the addition of photon number conservation \cite{dd}. If redshift remapping does not break any of these conditions then distance duality will still hold if remapping is treated correctly (although it may appear to be violated if remapping is not allowed for). However, distance duality may be fundamentally violated, e.g. if photon number is not conserved, providing a new test of redshift remapping that would yield different results than the ones discussed in this paper. As with many aspects of redshift remapping, this needs to be considered on a case-by-case basis. 

One aspect of this is that in general theories which exhibit remapping will also deform the value of the observables, e.g. inhomogeneous geometries will distort $d_L(z)$ and $d_A(z)$ due to lensing \cite{lens}, something that has to be considered self-consistently in any particular model with remapping.

\begin{table}[htbp]
   \centering
   \begin{tabular}{@{} lcccr @{}} 
   \toprule
      \multicolumn{2}{c}{} \\
     &  & Remapping & Remapping & $\Lambda$CDM\\
  Params &Prior & No-Planck & + Planck & No-Planck \\ 
   \midrule
   \hline
      $\Omega_m$ & [$10^{-2}$,1]      & $0.38\pm 0.06$ & $0.28\pm 0.02$ &  $0.32\pm 0.04$\\
        $H_0$             & [20,100]    & $71.1 \pm 4.2$ & $67.4\pm 2.7$ & $70.7 \pm 2.4$\\
      $\Omega_K$      & [-1,1]  & $-0.07\pm0.13$  & $-0.03\pm 0.03$ & $-0.17 \pm 0.09$\\
           $\alpha_1$     &[-10,10]      & $ 0.97\pm 0.06  $ & $1.04\pm 0.05$   &1 \\
      $\alpha_2$       & [-10,10]    & $0.05 \pm 0.03 $ &$0.00 \pm 0.02$ & 0 \\
      $\Omega_\Lambda$ & Derived & $0.68 \pm 0.15$ & $0.76 \pm 0.06$ & $0.85\pm 0.07$\\
         \hline
         \\
   \end{tabular}
   \caption{Marginalised constraints on parameters with and without redshift remapping for the combined $d_L(z), d_A(z)$ and $H(z)$ datasets (No-Planck). In the 4th column we show the results of including the Planck measurements of the CMB shift parameters $R$ and $l_a$, as described in section (\ref{otherdata}). Although this gives tighter constraints on the curvature, it is more model-dependent and does not significantly improve the constraints on the remapping parameters because we assume that the redshift of the surface of last scattering is unremapped.}
   \label{tab1}
\end{table}

\subsection{The impact of remapping data}
\label{remap}
As we have stressed, redshift remapping will change the inferred value of data points. Perhaps the most obvious is $H_0$ derived from local distance measurements and the Hubble law, which as shown by Eq. (\ref{taylor}), changes according to $H_0 \rightarrow H_0/\alpha_1$. If $\alpha_1 \neq 1$ there would be a disagreement between $H_0$ inferred this way and from those fitting global cosmological data such as the CMB or measurements using low-$\zo$ observations of $H(z)$ which need not show any change. This may provide an interesting alternative explanation worth further study for the current observed tension in $H_0$ \cite{SH03}.

The $H(z)$ values determined from cosmic chronometer data points \cite{cc} will also be rescaled.  Starting from $\dot{z} = -(1+z) H(z)$ and using $\dot{z} = z'\,\dot{z}_{\rm obs}$, we arrive at:
\begin{equation}
H(z) = -\frac{z' \dot{z}{}_{\rm obs}}{1 + z}
\label{ccrm}
\end{equation}
What happens to the BAO measurements? Apart from the change to the sound horizon that occurs when we explore regions of parameter space far from $\Omega_M = 0.3$, $h = 0.7$, the transverse BAO results only change implicitly, via the change in the value of $z$. However the radial BAO measurements, which yield $H(z)$, depend explicitly on $z$ \cite{baoreview}.

A minimal model for the effect proceeds as follows. Consider the flat FLRW line-element for radial geodesics: $dt = -dr/a$. Now $1+z = 1/a$ which yields the relation $dt = -dz/((1+z)H)$ and the usual result: $dr = dz/H(z)$. However, $dz$ is not observable from galaxy spectra, only $d\zo$ is. These two are related by $dz = z'~dz_{obs}$, which gives: 
\begin{equation}
H(z) = z' H_{\rm NR}(\zo)
\label{baorm}
\end{equation}
where $H_{\rm NR}(\zo)$ is the published value that arise from a BAO surveys with no remapping.  

How important are these changes on the data for our MCMC results? In Fig.(\ref{nos}) we compare the $\Omega_{\Lambda}-\alpha_1$ contours that arise from using remapped $d_A(z)$ and $H(z)$ data versus using the raw values from WiggleZ, BOSS and cosmic chronometers surveys without adjustment (to clarify, we do remap $z = f(\zo)$ in all cases). The impact on the $d_A(z)$ contour is fairly small. However the change on $H(z)$ is dramatic, and as a result, the combined contour is also significantly shifted and different in size, indicating that the changes redshift remapping has on the inferred data can be large. 

Note that this remapping of the $H(z)$ data largely breaks the degeneracy with remapping. This is partly due to the fact that the $H(z)$ data we use comes from different origins (BAO and cosmic chronometers) which scale differently under remapping, c.f. Eq's (\ref{ccrm}) and (\ref{baorm}). Somewhat like the Alcock-Paczynski test, the $H(z)$ data will only be mutually consistent when one has the correct remapping function. Thus, even measurements of just one observable -- $H(z)$ in this case -- can constrain remapping significantly if the measurements come from multiple probes which scale differently under remapping.

Redshift remapping can also alter error bars, even if the central value is unaffected. Consider the luminosity distance, $d_L(z)$, measured by Type Ia supernovae (SNIa), which are standardisable candles thanks to the Philips relation which encodes the fact that intrinsically brighter supernovae have wider lightcurves. However, this is degenerate with the time-dilation from the cosmological redshift,  which, if unknown, will lead to larger errors on SNIa distances. We do not, however, include this in our analysis due to the complexity of analysis it involves and because it will not fundamentally change our results. See \cite{margia} for derivations of the Hubble diagram without supernova redshifts. In addition, the peculiar redshift errors on the SNIa will be remapped in general, leading to further enhancement of the distance errors.

\subsection{Other data}\label{otherdata}

What are the implications of redshift remapping for other data? What about the Cosmic Microwave Background (CMB)? This is a difficult question to address in the model-independent spirit we have adopted in this paper which has looked at constraints on remapping functions of the form in Eq. (\ref{a1a2eq}) adapted mimicking the effects of acceleration at low-$\zo$ and which cannot be expected to hold at $z \sim 1100$. 

Of course, the CMB provides a wonderfully accurate measurement of quantities related to the sound horizon and the angular diameter distance to the surface of last scattering that are efficiently encoded in the CMB shift parameters \cite{Mukherjee2007}:
\begin{eqnarray}
R=\sqrt{\Omega_mH_0^2}r(z_\star)/c \\
l_a=\pi r(z_\star)/r_s(z_\star)
\end{eqnarray}
In the above equations $r(z_\star)$ and $l_a$ are the comoving sound horizon and angular size of the comoving sound horizon at the surface of last scattering, $z_\star$, respectively. In general we may expect models with redshift remapping to have non-trivial impact on $z_\star$. 

We have made two limited studies related to the CMB. Assuming that $z_\star$ is remapped via Eq. (\ref{a1a2eq}) we were unable to obtain convergent MCMC chains even with a prior of $|\alpha_2| < 10^{-3}$. This is not surprising since the $\zo^2$ term is ${\cal O}(10^6)$. We did not pursue this further since it is unreasonable to expect Eq. (\ref{a1a2eq}) to hold to decoupling and beyond. 

At the other extreme we studied the case in which $z_\star$ is unremapped corresponding to the case where remapping is unimportant before decoupling. This is appropriate for models in which remapping occurs because of the growth of inhomogeneity or backreaction at late times. In this case the CMB data points are independent of remapping and hence only improve the constraints on $\alpha_1, \alpha_2$ via breaking degeneracies with the standard cosmological parameters. 

\begin{figure}[ht!]
\centering
\includegraphics[width=\linewidth,height=7.5cm]{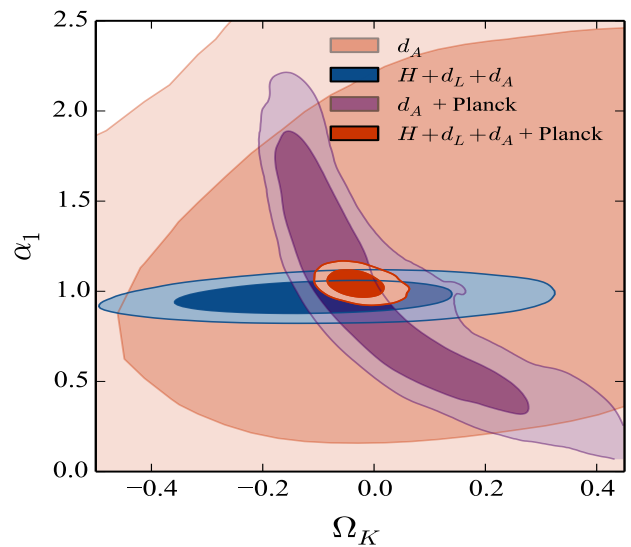}
\caption{Impact of including the Planck shift parameter data $l_a$ and $R$ into the $d_A(z)$ data bundle, assuming that the redshift of decoupling, $z_\star$ is unchanged by redshift remapping. As in the case with no remapping, the high-redshift CMB point allows the curvature to be measured much more precisely while giving only modest improvement in constraints on the remapping parameters; see Table (\ref{tab1}).} 
\label{planck}
\end{figure}

The derived mean values and covariance matrix of  $(l_a,R,\omega_b \equiv \Omega_b h^2)$ for the Planck data that we use in our MCMC analysis are given in \cite{Wang2013}. Adding these data points into our $d_A(z)$ basket significantly tightens constraints on the cosmic curvature as occurs in standard cosmology but has relatively little impact on the remapping parameters when all data is included, as shown in Fig.(\ref{planck}) and in Table (\ref{tab1}). 

We now briefly discuss other data. CMB lensing is sensitive to both the geometry of the universe and the growth factor, so will provide some constraints, but it is another integrated result, so may be relatively insensitive to remapping.  

Growth measurements require the change of the matter power spectrum over time together with measurements of the bias which can be obtained either via redshift-space distortions, which will be distorted in the same way we discussed for the radial BAO, $dz \rightarrow z'\,d\zo$, or via measurements of the bispectrum which will implicitly depend on distances between galaxies and hence will be deformed by remapping. Hence, although growth will likely provide strong complementary probes of remapping, the data will need to be consistently reanalysed.  

One strong measurement constraint on redshift remapping is the correlation between the CMB and large scale structure. In the standard model this is due to the Integrated Sachs Wolfe effect arising due to the decaying of gravitational potentials when the universe begins to accelerate. If there was significant remapping and no acceleration, how would the correlation arise? It would not within our simple formulation $z = f(\zo)$. However, a model such as the nonmetric gravity \cite{nonmetric1}, and perhaps even nonlinear electrodynamics \cite{nled}, may be able to create such a correlation since the remapped redshift is sensitive to ambient densities and magnetic fields in these models, which may cause interesting CMB-large scale structure correlations. 

An interesting way to constrain models with environment-sensitive redshift remapping is the comparison of BAO results from high-density probes like galaxies and low-density probes such as the Lyman-alpha forest and 21cm HI. These same kinds of tests should be powerful probes of other theories too, such as the Timescape cosmology and perhaps even Chameleon theories. The consistency of the BOSS Ly-$\alpha$ BAO result at $z=2.3$ with BAO results based on galaxies  suggest these models are already under pressure \cite{boss,boss2}. 

Finally, we note that since the strongest constraining power from $d_A(z)$ measurements come from beyond the turnover at $z \sim 1.5$, this is a new argument in favour of high-redshift BAO surveys such as those based on Lyman-$\alpha$ or Lyman break methods. 

\section{Case B}
\label{caseB}

{\em Case B} corresponds to the case where redshift remapping is assumed to affect redshifts as well as cosmic distances and the expansion rate. As a warm-up case, if we assume isotropy and homogeneity then irrespective of the theory of gravity we have the relation (see e.g. \cite{cop1}): 
\begin{equation}
d_L(z) = \frac{(1+z)}{H_0\sqrt{-\Omega_K}}\sin\left( \sqrt{-\Omega_K} \int \frac{dz}{E(z)}\right)\,.
\label{distance1}
\end{equation}
However unless we prescribe a theory of gravity we cannot link the cosmic dynamics (encoded in $E(z)$) to the matter content of the universe ($\Omega_m$, $\Omega_{\Lambda}$ etcÉ). 

How should one proceed in this case? One option is to expand $E(z)$ in a suitable set of basis functions or simply as a Taylor series. Since we wish to consider data at high redshifts  it is appropriate to use the CPL variable $\epsilon \equiv z/(1+z)$ which is always less than unity \cite{CPL, PS} in the expansion:   
\beq
H(\epsilon) =  H_0\left [1 + \sum_{n=1}^k \eta_n \epsilon^n \right] 
\label{expandH}
\eeq
where the $\eta_n$ are parameters that must be constrained with data, along with $H_0$. The choice of the maximum number of terms in this expansion, $k$, encodes the prior on how quickly one believes the Hubble rate can vary with redshift. Inserting eq. (\ref{expandH}) into eq. (\ref{distance1}) allows the Hubble diagram to be computed in terms of the coefficients $\eta_n$.  
Measurements of distances and Hubble rate can then be combined to constrain the $\eta_n$. Once a specific theory of gravity is assumed the $\eta_n$ would link to parameters related to the stress-energy tensor of space-time. 

However, while this is interesting, it is not the most general case: it still assumes FLRW geometry. Since inhomogeneous models exhibit interesting redshift remapping effects  \cite{shrink, backreact, CF, drwho, timescape} we would like to have an analysis that is as a general as possible. This is {\em Case B}:  eq. (\ref{distance1}) does not hold and we therefore do not know the link between $d_L(z)$ and $H(z)$. In addition we also have no Einstein-like equations for $H(z)$ itself. If we further assume that distance-duality does not hold, then there is no link between $d_L(z)$ and $d_A(z)$ and nothing can be said: we have data for the three functions $d_L(z), d_A(z), H(z)$ but we need to measure four functions (the previous three plus $z = f(\zo)$).

To make progress we must assume something. Perhaps the most appropriate place to start is the reciprocity relation \cite{eth}. If we further assume photon conservation this gives distance duality \cite{dd} and our desired constraint: $d_L(z) = (1+z)^2 d_A(z)$. This provides a fundamental relation that is likely to be true in most theories of gravity (although the photon conservation assumption may be wrong; see e.g. \cite{axionphoton}). 

Since we no longer have eq. (\ref{distance1}) we also expand $d_A(z)$ in its own independent Taylor series in powers of $\epsilon \equiv z/(1+z)$:
\beq
d_A(\epsilon) = \sum_{n=1}^k \delta_n \epsilon^n
\label{expandD}
\eeq
with free coefficients $\delta_n$ that must be constrained by data. Because we assume distance duality we can write $d_L(z)$ uniquely in terms of the $\delta_n$ once we have also specified the redshift remapping, which as before we take to be of the form:
\begin{equation}
z = \alpha_1\,\zo + \alpha_2\,\zo^2\,.
\label{zzo}
\end{equation}
To determine the optimal expansion order, $k$, we simulate $\Lambda$CDM data both with and without remapping and both for current and future data covariance matrices.  We then run MCMC reconstructions on the simulated data using different expansion orders $k=3, 4, 5, 6$, and with priors $[-10^{4}, 10^{4}]$ on $\eta_i$, $[10,100]$ on $H_0$, and $[100,200] Mpc$ on $r_s$. Although the $k = 3, 4$ runs fit some of the data well, the recovered $H_0$ values show a large bias. For the $k=6$ case on the other hand,  there was no bias but the errors on the redshift remapping parameters were very large, due to the enormously large parameter volume the MCMC can cover and give a reasonable fit to the data \cite{PS}.  We find the $k=5$ expansion to be a good compromise between unbiased estimates of $H_0$ and $r_s$  and the tightest constraints on the redshift remapping parameters.  We therefore chose this case in all our analysis of {\em Case B} which, along with $\alpha_1, \alpha_2$ and $H_0$, brings the number of free parameters that must be estimated from the data  to 13. In analysing real, high-quality, future data there is of course no assurance that the curves will look like $\Lambda$CDM and hence $k = 5$ should not be assumed. Rather, a model-selection method such as the Bayesian evidence should be assumed in order to select the optimal value of $k$ from the data \cite{PS}. However in our present analysis this is an irrelevant technical detail.  Note that one must ensure that the priors on the parameters are chosen broad enough not to artificially introduce bias in the recovered parameters - the best-fit  parameters $\eta_n, \delta_n$ grow rapidly with increasing $k$ as the terms with alternating powers of $\epsilon$ need to balance to higher and higher relative precision.

Despite the very wide range of freedom the 13 parameters imply we can make significant progress by noting that if we have separate measurements of $d_A(z)$ and $d_L(z)$ at the same redshifts (e.g. from SNIa, BAO or lensing), then there is no residual freedom other than that encoded in redshift remapping (since  $d_L(z)/d_A(z) =  (1+z)^2$) and hence we can constrain remapping completely independently of the parameters $\eta_n, \delta_n$. This also shows that we can successfully consider much more complex expansions than eq. (\ref{zzo}). Indeed, with perfect $d_L$ and $d_A$ data one could reconstruct $z = f(\zo)$ perfectly without knowing anything about the theory of gravity or the geometry of the universe.
In reality it is rare to have measurements of $d_L$ and $d_A$ at exactly the same redshift and hence in practise there is some covariance between the parameters $\alpha_{1,2}$ and $\eta_n, \delta_n$ which is mitigated by getting more data. 

To determine with the $H(z)$ curves show acceleration is non-trivial. One might be tempted to ask what fraction of the MCMC curves had negative $q_0$. However, this is almost meaningless in this case. Because they are constructed via the Taylor-series expansions (\ref{expandH}) and (\ref{expandD}) they are in general very badly behaved outside the range where there is significant data, the so-called Runge phenomenon \cite{runge}. In particular the value of $q_0$ oscillates wildly since there is no data at $z = 0$ and $q_0$ involves two derivatives. This can somewhat be seen in Fig. (\ref{dota1}) where there are many example curves of both positive and negative $q_0$.  

As a result we focussed on the redshift region $0.2 < z < 0.8$ where there is significant amounts of data. We checked that main conclusions are not sensitive to changing both the upper and lower range of this interval, as long as there is data. We search for acceleration by looking for an increasing $\dot{a} \equiv H(z)/(1+z)$ for decreasing $z$ (i.e. $\ddot{a} > 0$) for at least one redshift in the range $0.2 < z < 0.8$. This does not imply that there will be acceleration across the whole interval - many of the curves show both acceleration and deceleration in the range. However, our main interest is whether acceleration can be completely avoided using redshift remapping, which requires that there be no accelerating phases at all.  

We have used this approach with the current SNIa, BAO and cosmic chronometer data and find that distance duality strongly constrains all the parameters to the extent that $99\%$ of all MCMC chain points exhibit acceleration somewhere in the region $0.2 < z < 0.8$.   This shows that acceleration is a generic feature of models fit to the current data even when we assume neither General Relativity nor FLRW geometry. This is consistent with expectations from $\Lambda$CDM. We simulated fake data from $\Lambda$CDM models with no remapping using the current data covariance matrix, finding that $99.5\%$ of the chain points showed acceleration for at least one redshift in the region $0.2 < z < 0.8$. This shows that the degeneracy between acceleration and remapping is broken and provides strong supporting evidence that the current data is consistent with no remapping. We note that this conclusion can probably be evaded by considering more complex models of remapping with more parameters exploiting the lack of overlapping $d_A$ and $d_L$ data in certain redshift regions, but we have no {\em a priori} reason to consider such models.

\begin{figure}[ht!]
\centering
\includegraphics[width=8.5cm]{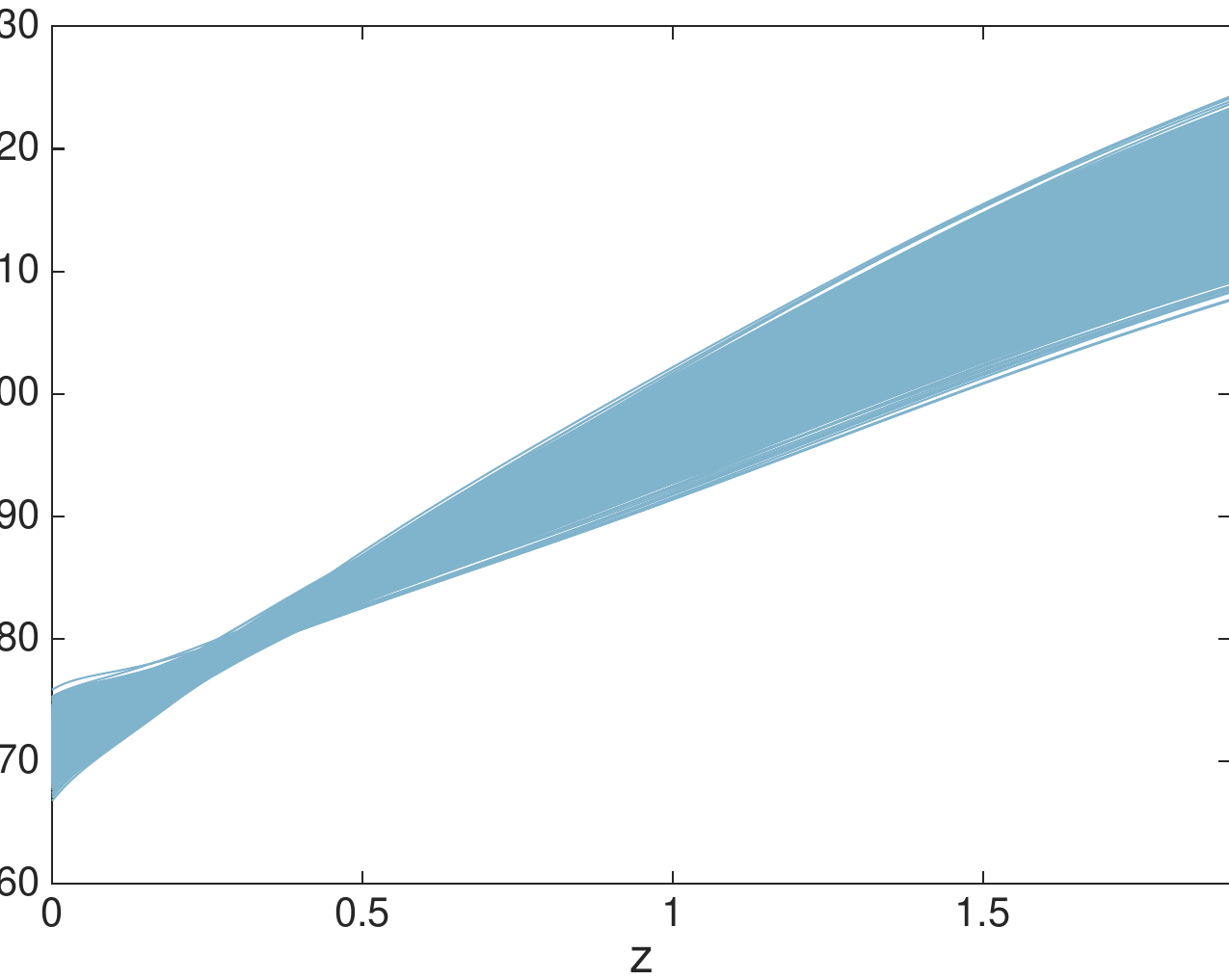}
\caption{Future data MCMC reconstructions of $H(z)/(1+z)$ from a fiducial Einstein-de Sitter (EdS) non-accelerating model: none of the chains exhibit any acceleration showing that future data will be able to rule out redshift remapping. In contrast, simulations of an EdS model using current data covariance matrices show that about $54\%$ of chain elements have some (spurious) acceleration, due to the gaps in the data and the Runge phenomenon. Hence we cannot yet fully exclude remapping as an alternative to acceleration. Compare with Fig. (\ref{dota1}).} 
\label{EdS_Future}
\end{figure}

As a final test we simulated an Einstein-de Sitter (EdS) model with no acceleration and no remapping and tested the chains for acceleration. We found that approximately $54\%$ of the chains exhibited some spurious acceleration when fit to non-accelerating data, due to the large gaps in current data, the intrinsically oscillatory nature of expansions and the Runge phenomenon. It would be interesting to consider alternative general methods, such as Pade' approximants, that may lead to fewer spurious oscillations in the reconstructed $H(z)$ but this is left to future work. The oscillations imply that we cannot yet conclusively rule out redshift remapping as an alternative to acceleration, although it is rather disfavoured. Future data will allow us to make significant improvements however. In Fig. (\ref{EdS_Future}) we show the reconstructed $H(z)/(1+z)$ curves from the MCMC fits to the future data for a fiducial EdS model. None of the chains have acceleration due to the dense redshift space sampling of the data. This shows that within a decade or so we will be able to conclusively rule out redshift remappings of the sort considered in this paper as an alternative to cosmic acceleration. 
\end{document}